\documentclass[showkeys,twocolumn,showpacs,preprintnumbers,amsmath,amssymb]{revtex4}
\usepackage[pdftex]{graphicx}
\usepackage{color} 

\pdfoutput=1
\begin{document}
\title{Asymmetric field dependence of magnetoresistance in magnetic films.}

\author{A. Segal}
\email{amirsega@post.tau.ac.il}
\homepage{http://star.tau.ac.il/~gnl}
\author{O. Shaya}
\author{M. Karpovski}
\author{A. Gerber}
\affiliation{Raymond and Beverly Sackler Faculty of Exact Sciences, School of Physics and Astronomy, Tel Aviv University, Ramat Aviv 69978, Tel Aviv, Israel}

\date{\today}

\begin{abstract}
We study an asymmetric in field magnetoresistance that is frequently observed in magnetic films and, in particular, the odd longitudinal voltage peaks that appear during magnetization reversal in ferromagnetic films, with out-of-plane magnetic anisotropy. We argue that the anomalous signals result from small variation of magnetization and Hall resistivity along the sample. Experimental data can be well described by a simple circuit model, the latter being supported by analytic and numerical calculations of current and electric field distribution in films with a gradual variation of the magnetization and Hall resistance. 

\end{abstract}

\maketitle

Onsager's reciprocity relations \cite{Onsager} are the cornerstone in understanding the field symmetry of magnetotransport measurements. Magnetoresistance or longitudinal resistivity (measured along the current flow direction) is predicted to be an even function of magnetic induction B, while transverse (Hall) resistivity is specified to be odd with respect to B when a magnetic field is applied perpendicular to the sample plane. General acceptance of these rules is so common that in numerous experimental cases, when current and voltage contacts can not be arranged in a well defined 5-probe geometry, the magnetoresistance and Hall effect data are respectively extracted as the even and odd in field components of the measured 4-probe signal. However, asymmetric in field magnetoresistance is quite frequently observed in magnetic materials (also in samples with fully symmetric magnetic properties and properly arranged current and voltage contacts) \cite{two,three}, although this is rarely mentioned and discussed \cite{four,five}. A seeming violation of Onsager's law only recently attracted attention when sharp distinctive peaks of magnetoresistance, odd with respect to applied field, were found at magnetization reversal of ferromagnets with an out-of plane magnetic anisotropy \cite{Cheng,seven}. As argued by Cheng et al \cite{Cheng} the effect can appear when a domain wall (DW), located between the voltage probes, runs perpendicular to both magnetization and current. Electric fields generated by the extraordinary Hall effect (EHE) have opposite polarities on both sides of the DW, which can produce a circulating current loop and a respective extra voltage contribution. The model was used to explain the odd in field longitudinal voltage peaks in a specially designed Co-Pt multilayer film with a single DW gradually propagating along the sample. However, the effect was also observed in other samples with multiple domains \cite{Cheng,eight}, and the applicability of the "single wall" model in this general case is dubious. 

   In this paper we present two typical cases of asymmetric magnetoresistance observed in magnetic films and analyse their origins. We shall argue that the anomalous behaviour can consistently be explained by a gradual variation of magnetization and Hall resistivity along the sample. 

\begin{figure}
\includegraphics [width=9cm,height=9cm]{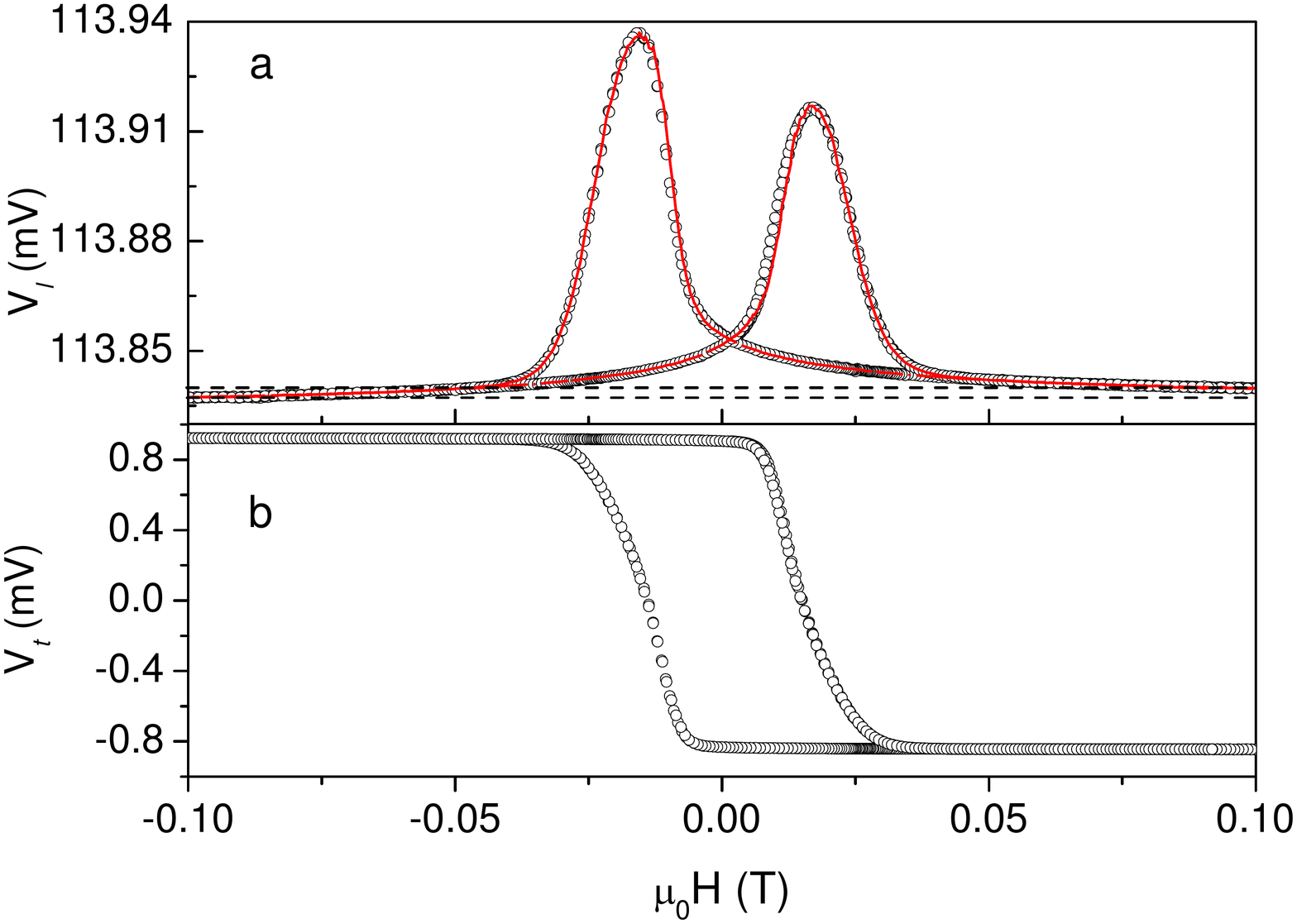}
\caption{\label{fig:one} (a) Longitudinal voltage ($\circ$) measured in a 6 nm thick Ni film at 4.2 K as a function of field applied perpendicular to the film plane. Solid line (\textcolor{red}{--}) is a fit calculated according to Eq. \ref{eq:eight}. Dashed lines are a guide for the eye that emphasizes the high field asymmetry. (b) Hall voltage ($\circ$) measured simultaneously.}
\end{figure}

   Fig. \ref{fig:one}a presents the longitudinal voltage $V_l$ measured in a 6 nm thick Ni film at 4.2 K as a function of a magnetic field applied perpendicular to the film plane in both field polarities. Thin Ni films possess the surface induced out-of-plane anisotropy at low temperatures \cite{Riss} responsible for the hysteresis in the magnetoresistance curve. Anisotropic magnetoresistance is the origin of the negative magnetoresistance when a field is applied perpendicular to the electric current direction. $V_l$ reaches maximum at magnetic fields corresponding to the coercive field value when the macroscopic out of plane magnetization crosses zero. Notably, the magnitude of the maximal voltage is not equal at two field polarities although the location of the peaks is the same.  Similar asymmetric maxima can be found in several publications \cite{ten,eleven,twelve}. The measured voltage is Ohmic (linear in electric current) and, if interpreted as magnetoresistance, its asymmetry would mean a violation of the Onsager rule. 

   Fig. \ref{fig:one}b presents the transverse (Hall) voltage for the same sample. Hall voltage in magnetic films depends on magnetization as \cite{Hurd}: 

\begin{equation}
\label{eq:one}
V_t=\frac{I}{t}(R_0B+\mu_0R_{EHE}M)
\end{equation}

where $I$ is electrical current, $t$ - thickness, $R_0$ and $R_{EHE}$ are the ordinary and extraordinary Hall coefficients, $B$ and $M$ are the out of plane components of magnetic field induction, and magnetization respectively. A clear hysteresis loop is seen in Fig. \ref{fig:one}b which is proportional to magnetization (the EHE term contribution is much larger than the ordinary one; therefore we neglect the ordinary Hall component in the following discussion).

\begin{figure}
\includegraphics [width=8cm]{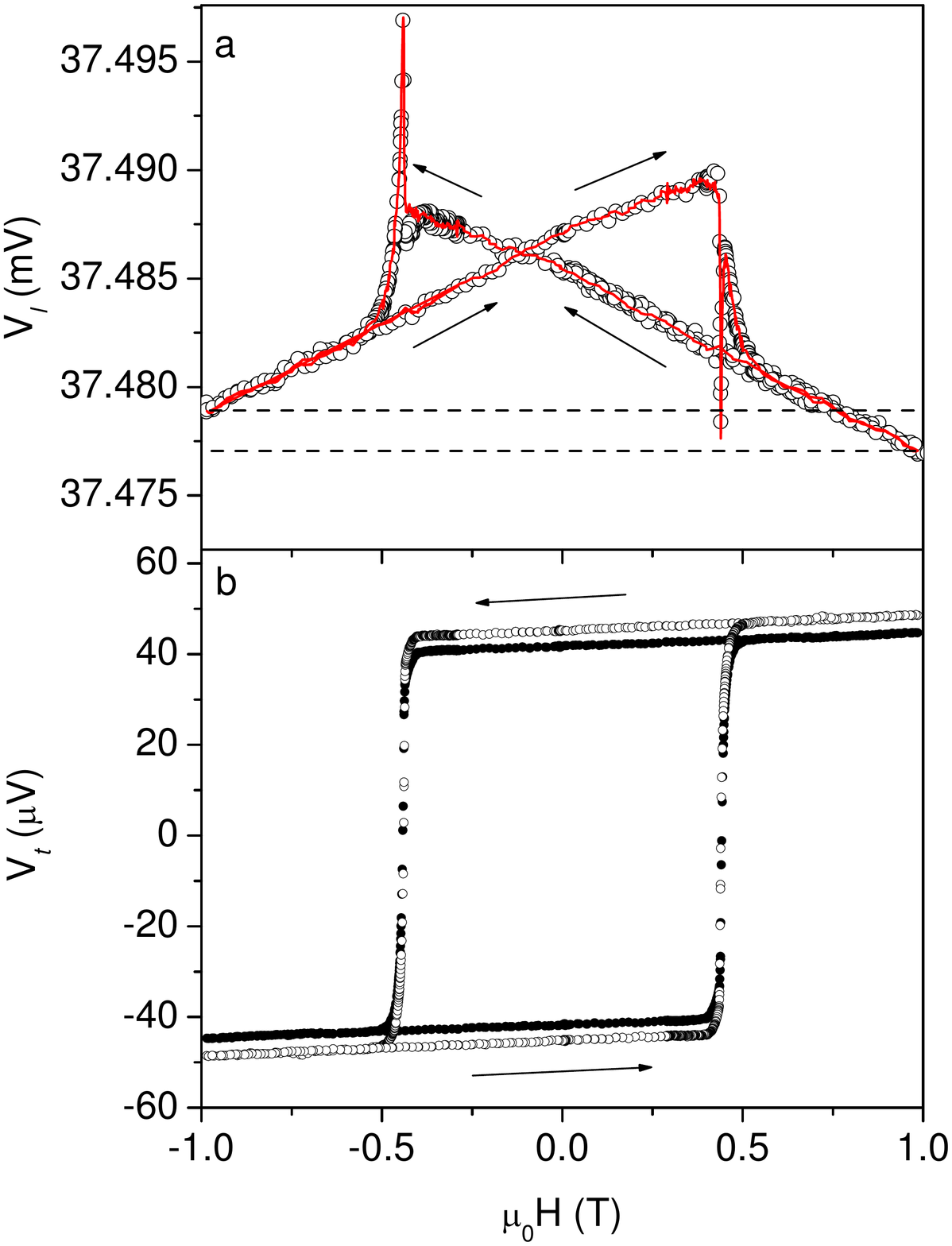}
\caption{\label{fig:two}(a) Longitudinal voltage $V_l$ ($\circ$) measured in the Co/Pd multilayer sample at 4.2 K as a function of applied field normal to the film plane. Solid line (\textcolor{red}{--}) is a fit according to Eq. \ref{eq:eight}. Dashed lines are a guide for the eye that emphasizes the high field asymmetry. Arrows indicate the direction of the field sweep. (b) Hall voltages $V_{t1}$ ($\circ$) and $V_{t2}$ ($\bullet$) measured simultaneously at two locations along the sample. }
\end{figure}

   Another striking example of asymmetric in field magneto-voltage is presented in Fig. \ref{fig:two}a. The longitudinal voltage measured in a Co/Pd multilayer sample (10 bilayers of 0.2 nm thick Co and 1.1 nm thick Pd, total thickness 13 nm) is shown as a function of a magnetic field normal to the film. The sample was prepared by sequential e-beam deposition of Co and Pd layers on a GaAs substrate. It has the six-contacts Hall bar geometry, 5 mm wide and 15 mm long. The distance between longitudinal and transverse voltage contacts is 5 mm. The sample has a strong out-of-plane anisotropy, typical for Co/Pd multilayers. Sharp antisymmetric peaks are clearly observed at about 0.44 T when magnetization reverses its polarity. The antisymmetric peaks are superimposed with a slightly asymmetric magnetoresistance curve. It is important to note that the polarity of the odd peaks (positive in the negative field and negative in the positive field) is reversed if measurement of the longitudinal voltage is done along the opposite edge of the film.  A similar effect was found by us in FeTb films \cite{Shaya} and was previously reported in Co/Pt multilayers \cite{Cheng} and (Ga,Mn)As epilayers \cite{seven} with perpendicular magnetic anisotropy. Following Cheng et al \cite{Cheng} the odd in field longitudinal voltage signal can appear when a domain wall separating two domains with up and down magnetization is located between the voltage probes. Electric fields generated by the EHE depend on the local magnetization and have opposite polarities on both sides of the domain wall. These electric fields normal to the current can produce a circulating current loop around the domain wall and the respective additional voltage contribution along the sample. The model of a single domain wall assumes that magnetization is opposite at locations of the two longitudinal voltage contacts when the anomalous voltage peaks appear. This assumption can be tested experimentally by measuring the Hall voltage at two cross-sections along the sample. Fig. \ref{fig:two}b shows $V_t$ measured between two pairs of contacts transversal to the current direction, at two locations along the sample ($\circ$ and $\bullet$), while the longitudinal voltage $V_l$, shown in Fig. \ref{fig:two}a, is measured simultaneously between a pair of longitudinal contacts. The magnetization reverses almost simultaneously at both locations (the difference in coercive fields is about 20 Oe, whereas the reversal width defined as the field span over which $V_t$ varies between 10\% and 90\% is approximately 700 Oe (Fig. \ref{fig:two}b)). The antisymmetric longitudinal voltage peaks (Fig. \ref{fig:two}a) appear with the reversal of magnetization. The width of the peaks is equal to the width of the magnetization (Hall voltage) reversal. This observation does not agree with the "single domain wall" picture that predicts opposite Hall voltage polarities at two cross-sections when the anomalous peaks appear. Two more experimental results are important for future discussion: (i) $V_t$ signals at two cross-sections are similar in shape but differ in magnitude in the magnetically saturated state at high fields by approximately 8\%, and (ii) macroscopic magnetization is not uniform: there is a small but finite difference in coercive fields along the sample.

\begin{figure}
\includegraphics [width=4cm]{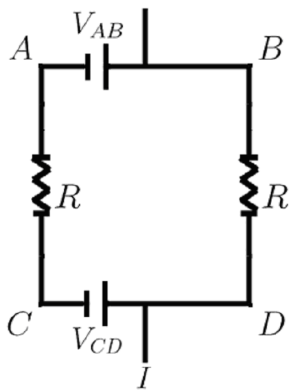}
\caption{\label{fig:three}Effective circuit representation of the sample.}
\end{figure}

   Although the single wall model is not in agreement with the experimental data, one can assume that the transverse voltage is not uniform along the sample. We, therefore, model the sample as a simple circuit shown in Fig. \ref{fig:three}. $V_{AB}(H)$ and $V_{CD}(H)$ represent the transverse voltage generated by Hall effects at two cross-sections AB and CD, while two equal resistors $R$ are positioned between A and C, and B and D. Field dependence of the resistors $R(H)$ is the usual symmetric in magnetic induction magnetoresistance.  Following Kirchhoff's circuit laws, the longitudinal voltages at two edges of the sample are:

\begin{equation}
\label{eq:two}
\begin{array}{cc}
V_{AC}=\frac{IR+V_{AB}-V_{CD}}{2} 	\\
V_{BD}=\frac{IR-V_{AB}+V_{CD}}{2}
\end{array}
\end{equation}

Voltage measured along the sample would differ from the ordinary Ohmic   if $(V_{AB} - V_{CD}) \neq 0$, i.e. $V_{AB} \neq V_{CD}$. The field symmetry of $(V_{AB} - V_{CD})$ is even in the case of e.g. non-uniform planar Hall effect contribution \cite{fifteen} or odd when the ordinary and / or extraordinary Hall effects are present. In this case $(V_{AB} - V_{CD})$ is given by:

\begin{equation}
\label{eq:three}
V_{AB}-V_{CD}=\mu_0\cdot I \left ( \frac{R_{EHE,AB}M_{AB}}{t_{AB}}-\frac{R_{EHE,CD}M_{CD}}{t_{CD}}  \right )
\end{equation}

with $t_{AB}$ and $t_{CD}$ as the local thickness,  $M_{AB}$ and $M_{CD}$ as the local magnetizations and $R_{EHE,AB}$ and $R_{EHE,CD}$  are the EHE coefficients at cross-sections AB and CD respectively.

   Several mechanisms can cause a non-uniform transverse voltage along the sample. The simplest is a gradual variation of thickness $t$ (\ref{eq:one}) due to either unintended imperfection of fabrication or when wedge samples are studied. This argument is applicable to any material including non-magnetic metals \cite{sixteen} and semiconductors \cite{seventeen}. In magnetic materials there are additional mechanisms that can affect the EHE coefficient $R_{EHE}$. In thin ferromagnetic films $R_{EHE}$ depends on the thickness and diverges in the thin film limit due to an enhanced surface scattering \cite{three,eighteen}. In granular ferromagnetic or superparamagnetic films $R_{EHE}$ depends on size, density and shape of magnetic clusters that might not be uniform along the sample, due to deposition and annealing procedures \cite{four}. If magnetization is uniform along the sample ($M_{AB}=M_{CD}$), Eq. \ref{eq:three}  gives:

\begin{equation}
\label{eq:four}
V_{AB}-V_{CD}=\left (  1-\frac{R_{EHE,CD}\cdot t_{AB}}{R_{EHE,AB}\cdot t_{CD}} \right )\cdot V_{AB}
\end{equation}

The longitudinal voltage $V_{BD}$ can then be presented as:

\begin{equation}
\label{eq:five}
V_{BD}=\frac{IR}{2}-\alpha \cdot V_{AB}
\end{equation}

where  $\alpha$ is a coefficient that depends on thickness and $R_{EHE}$ variation along the sample. The first term on the right hand side of Eq. \ref{eq:five} is even with respect to the field while the second is odd and proportional to the transverse voltage. $V_{AB}$ is a monotonic function of field (see Fig. \ref{fig:two}b), therefore Eq. \ref{eq:five} can explain the high field asymmetry of the longitudinal voltage in Fig. \ref{fig:two}a, but not the antisymmetric peaks at the magnetization reversal. We then assume that magnetization is not uniform and reverses gradually along the sample with raise of the applied field. The local magnetization values $M_{AB}(H)$ and $M_{CD}(H)$ are connected by:
\begin{equation}
\label{eq:six}
M_{CD}(H)=M_{AB}(H)-\Delta H \cdot \frac{\partial M_{AB}(H)}{\partial H}
\end{equation}

where $\Delta H$ is the increase of applied field needed to propagate the magnetization reversal from the cross-section AB to CD. Then, to the first order of $\Delta H$:

\begin{equation}
\label{eq:seven}
\begin{array}{cc}
V_{AC}=\frac {1}{2}\left ( IR(H)+\frac {\Delta H \cdot I \cdot \mu _0 \cdot R_{EHE,AB}}{t} \cdot \frac{\partial M_{AB}(H)}{\partial H} \right ) \\
V_{BD}=\frac {1}{2}\left ( IR(H)-\frac {\Delta H \cdot I \cdot \mu _0 \cdot R_{EHE,AB}}{t} \cdot \frac{\partial M_{AB}(H)}{\partial H} \right )
\end{array}
\end{equation}

where we assume that $t_{AB}=t_{CD}\equiv t$ and $R_{EHE,AB}= R_{EHE,CD}$. The second term in Eq. \ref{eq:seven} is odd with respect to the field ( $H$ is odd) and can be significant in materials with a large EHE coefficient and sharp reversal of magnetization, as in thin ferromagnetic films with the out-of-plane anisotropy. The shape of $\Delta H \cdot \left ( \frac{\partial M_{AB}(H)}{\partial H} \right )$  has a strong peak at magnetization reversal; therefore, this term can account for the antisymmetric peaks, as in Fig. \ref{fig:two}a, or for a significant difference in the maximal resistance in Fig. \ref{fig:one}a. If both the gradual reversal of magnetization and the variation of the saturated high field Hall voltage along the sample are considered, the combination of Eqs. \ref{eq:one}, \ref{eq:five} and \ref{eq:seven} gives:

\begin{equation}
\label{eq:eight}
V_{BD}=\frac{1}{2}\left ( IR(H)-\Delta H \cdot \frac{\partial V_{AB}(H)}{\partial H}\right )-\alpha \cdot V_{AB}(H)
\end{equation}

We applied Eq. \ref{eq:eight} to fit the experimental data both for the Ni film (Fig. \ref{fig:one}a) and Co/Pd multilayer (Fig. \ref{fig:two}a) by using the measured transverse voltage $V_{AB}(H)$ and two fitting parameters   and  $H$. The asymmetric magnetoresistance of Ni (solid line in Fig. \ref{fig:one}a) was calculated with $\Delta H = 2.2 Oe$ and $\alpha = 1.4 \cdot 10^{-3}$. The fit for the Co/Pd multilayer, shown in Fig. \ref{fig:two}a by a solid line, was calculated with $\alpha  = 2\cdot 10^{-2}$ and $\Delta H = 24 Oe$. This value of $\Delta H$ is in good agreement with the measured 22 Oe difference in coercive fields between cross-sections AB and CD. It should be noted that only minor inhomogeneity ($\alpha$) and non-uniformity of magnetization reversal ($\Delta H$) along the sample are sufficient to generate large anomalous signals. A possible cause for variation of the coercive field along the sample is the thickness variation. Magnetization reversal was reported \cite{Cheng,nineteen} to propagate along wedge shaped samples with thickness variation of a few percents only. Other possible causes are variation of surface roughness and adhesion to the substrate which are suspected \cite{twenty} of inducing a transition from nucleation dominated reversal to domain-wall-motion reversal.

   Although the model presented above is in a good agreement with the experimental data, one can wonder if a simple circuit (Fig. \ref{fig:three}), which has only two current channels, provide a reliable description of a macroscopic sample. In the following we present a more rigorous derivation of the electric potential along an infinitely long sample with variable thickness and Hall resistivity, and show that in the proper limit the result is identical to Eq. \ref{eq:eight}. In order to reduce the problem to two dimensions, we follow Ref. 16 and define the following two-dimensional fields:

\begin{equation}
\label{eq:nine}
<\vec{j}(x,y)>\equiv \frac{1}{t}\int^{t(x,y)}_{0}\vec{j}(x,y,z')dz' 
\end{equation}

\begin{equation}
\label{eq:ten}
<\vec{E}(x,y)>\equiv \frac{1}{t(x,y)}\int^{t(x,y)}_{0}\vec{E}(x,y,z')dz' 
\end{equation}

where $\vec{E}$ is electric field, $\vec{j}$  - the current density, $t$ -  the average sample thickness and $t(x,y)$ is the actual sample thickness at each point. The two dimensional current distribution is determined by:

\begin{equation}
\label{eq:eleven}
<\vec{E}(x,y)>=\frac{t \cdot \overleftrightarrow{\rho}(x,y)}{t(x,y)} \cdot <\vec{j} (x,y)>
\end{equation}

\begin{equation}
\label{eq:twelve}
\vec{\nabla} \cdot <\vec{j}(x,y)>=0
\end{equation}

and to a good approximation by:

\begin{equation}
\label{eq:thirteen}
\vec{\nabla} \times <\vec{E}(x,y)>=0
\end{equation}

where $\overleftrightarrow{\rho}$ is the spatially dependent resistivity tensor. Boundary conditions are set to prevent current flow normal to the sample edges. Exponential variation of thickness along the sample with constant Hall resistivity was analysed by Bruls et al \cite{sixteen}. The equations were found to be identical to those describing a sample with an exponential variation of charge carrier density \cite{seventeen}. Adaptation of the latter case gives the field dependent potential along the sample as:

\begin{equation}
\label{eq:fourteen}
\varphi (x,y,H)=\frac{-\rho (H) \cdot I \cdot \beta (H)\cdot exp\left( \frac{-x+\beta(H)\cdot y}{a}\right)}{t \cdot \left[ 1-exp\left ( \frac{\beta (H)\cdot w}{a} \right)\right ]}
\end{equation}

where $\rho (H)$ is resistivity, $a$ - the length scale over which the sample thickness changes by a factor of e, $w$ - the sample width, and $ \beta (H)$ - the ratio of Hall and longitudinal resistivities. In the limit of $a>>x$ and $a>>\beta (H)·w$, Eq. \ref{eq:fourteen} can be reduced to:

\begin{equation}
\label{eq:fifteen}
\begin{array}{cc}
\varphi (x,y,H)=-\frac{\rho (H) \cdot I}{t \cdot w}[x+\beta (H)\cdot (w/2-y)] \\
+\frac{\rho (H)\cdot I\cdot x^2}{2\cdot t \cdot w \cdot a}+\frac{\rho (H) \cdot I \cdot \beta(H) \cdot x \cdot(w/2-y)}{t \cdot w \cdot a}
\end{array}
\end{equation}

where  $a=\frac{L \cdot t}{\Delta t}$,  with $L$ being the distance between the longitudinal voltage probes and $\Delta t$ the change of sample thickness between location of the longitudinal probes (at $x = L/2$). The first term in Eq. \ref{eq:fifteen} consists of the standard longitudinal and transverse voltages of a homogeneous sample. The second and third terms are corrections to the potential due to the thickness variation. The second term does not contribute to longitudinal voltage since it is symmetric in $x$. The third term is proportional to the Hall voltage and changes sign depending on the location of the probes (at $y = 0$ or $y = w$). 

   Linear variation of Hall resistivity due to change of charge carrier density along the sample was analysed by Ilan et al \cite{twentyone} in 2-D electron gas. In the case of magnetic materials we ascribe the gradient of Hall resistivity to linear variation of both magnetization and $R_{EHE}(t)$ along the sample, so that:

\begin{equation}
\label{eq:sixteen}
\begin{array}{cc}
\frac{\partial \rho_{xy}(x,H)}{\partial x}=\mu _0 \cdot M(0,H)\cdot \frac{\partial R_{EHE}(t)}{\partial t} \cdot \frac{\Delta t}{L} \\
+\mu _0 \cdot R_{EHE}(0) \cdot \frac{\partial M(x,H)}{\partial x}
\end{array}
\end{equation}

where $M(0,H)$ and $R_{EHE}(0)$ are the values of magnetization and $R_{EHE}$ at $x=0$. For  $\rho _{xy}$ varying along the $x$ coordinate only, and $l_x>>w$, the potential along the sample is given by:

\begin{equation}
\label{eq:seventeen}
\begin{array}{cc}
\varphi (x,y,H)=-\frac{\rho (H) \cdot I}{t \cdot w}[x+\beta (H)\cdot (w/2-y)] \\
-\frac{\rho (H) \cdot I \cdot x\cdot (w/2-y)}{t \cdot w \cdot l_x(H)}
\end{array}
\end{equation}

where $l_x(H)=\rho(H)\cdot \left( \frac{\partial \rho _{xy} (x,H)}{\partial x} \right)^{-1}$\cite{twentyone}.  The first term in Eq. \ref{eq:seventeen} corresponds to the potential distribution in a homogeneous sample, whereas the second term is the correction due to a spatial variation of the Hall resistivity. Since the correction terms in Eqs. \ref{eq:fifteen} and \ref{eq:seventeen} are small and of different origins, they are additive (higher order corrections are neglected). The longitudinal voltage can be calculated from Eqs. \ref{eq:fifteen}, \ref{eq:seventeen} as:

\begin{equation}
\label{eq:eighteen}
\begin{array}{ccc}
V_l(H)=\varphi (x=-L/2,H)-\varphi (x=L/2,H)=\frac{\rho (H)\cdot I \cdot L}{t \cdot w} \\
\pm\frac{I \cdot \mu _0 \cdot M(0,H)\cdot R_{EHE}(0)}{2\cdot t}\left ( \frac{\Delta t}{R_{EHE}(0)}\cdot \frac{\partial R_{EHE}(t)}{\partial t}-\frac{\Delta t}{t} \right ) \\
\pm \frac{L\cdot I \cdot \mu _0 \cdot R_{EHE}(0)}{2 \cdot t}\cdot \frac{\partial M(x,H)}{\partial x}
\end{array}
\end{equation}

where the $\pm$ sign stands for $y = 0$ (+) and $y = w$ (-). The first term in Eq. \ref{eq:eighteen} is simply $I·R(H)$, the second term is the correction due to the thickness variation and is proportional to the transverse voltage, and the last term is the correction due to a non-uniform magnetization along the sample. Finally, by assuming a constant ratio between the change of an applied field $\Delta H$ and the propagation of the magnetization reversal over a distance $L$, we calculate:

\begin{equation}
\label{eq:nineteen}
\begin{array}{cc}
V_l(H)=I\cdot R(H) \pm \frac{1}{2}\left ( \frac{\Delta t}{R_{EHE}(0)}\cdot \frac{\partial R_{EHE}(t)}{\partial t}-\frac{\Delta t}{t} \right )V_t(0,H) \\
\pm \frac{\Delta H}{2}\cdot \frac{\partial V_t(0,H)}{\partial H}
\end{array}
\end{equation}

with $V_t(0,H)=\frac{I\cdot \mu _0 \cdot R_{EHE}(0)\cdot M(0,H)}{t}$ . Eq. \ref{eq:nineteen} is identical to Eq. \ref{eq:eight} obtained from the circuit model.

   The analytic calculation was done for an infinitely long sample. In order to treat a finite sample, numerical calculations were carried out. Following Hajjar et al \cite{twentytwo}, the current distribution was calculated by taking the finite difference version of Eqs. \ref{eq:twelve} - \ref{eq:thirteen} on a two dimensional rectangular lattice. Boundary conditions were added along the length of the sample edges together with the current source and drain. The sample dimensions were chosen equal to the actual geometry of the Co/Pd multilayer film, in which $L = w$ and the total sample length is $3w$. The effective resistivity tensor that was used included both thickness and magnetization gradients along the sample:
   
\begin{equation}
\label{eq:twenty}
\frac{t\cdot \rho (x,H)}{t(x)}=\frac{\rho (H)}{1+x\cdot \frac{\Delta t}{t\cdot L}}\cdot \left ( \begin{array}{cc}
1 & \beta (H)+\frac{x}{l_x(H)} \\
-\beta (H)-\frac{x}{l_x(H)} & 1 \end{array} \right )
\end{equation}

In order to obtain $l_x(H)$ at each field value, the normalized change in magnetization between locations AB and CD was estimated as:   
\begin{equation}
\label{eq:twentyone}
\frac{\Delta M(H)}{M_S}=\frac{V_{CD}(H)}{V_{CD,S}}-\frac{V_{AB}(H)}{V_{AB,S}}
\end{equation}

\begin{figure}
\includegraphics [width=8cm]{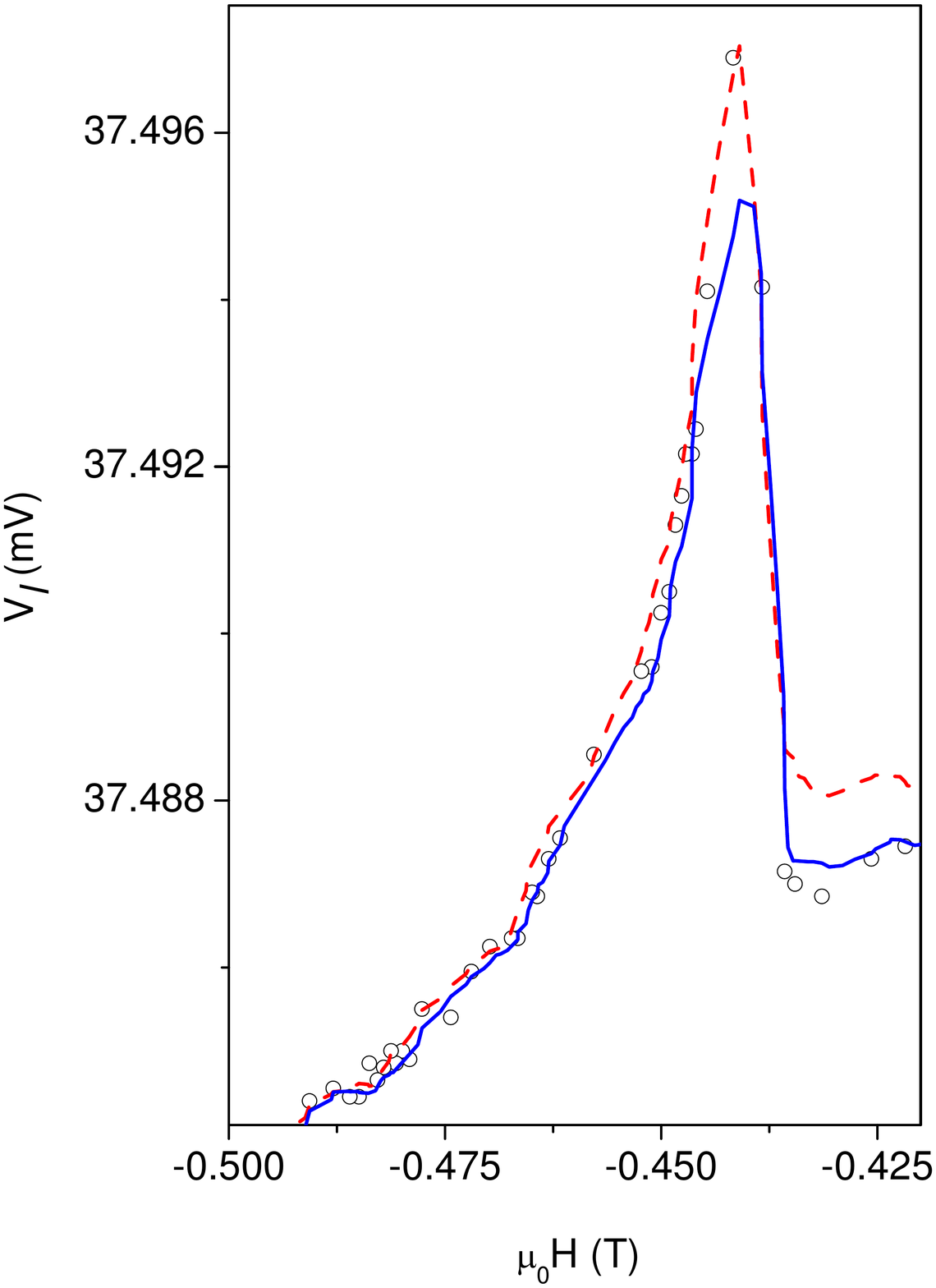}
\caption{\label{fig:four} Longitudinal voltage $V_l$ ($\circ$) measured in the Co/Pd multilayer sample at 4.2 K as a function of field normal to the film plane in the peak region. Solid line (\textcolor{blue}{--}) is a fit calculated by using Eqs. \ref{eq:twenty}, \ref{eq:twentyone}, dashed line (\textcolor{red}{- -}) is a fit according to Eq. \ref{eq:eight}.}
\end{figure}
\begin{figure}
\includegraphics* [width=8cm,viewport=50 150 500 720]{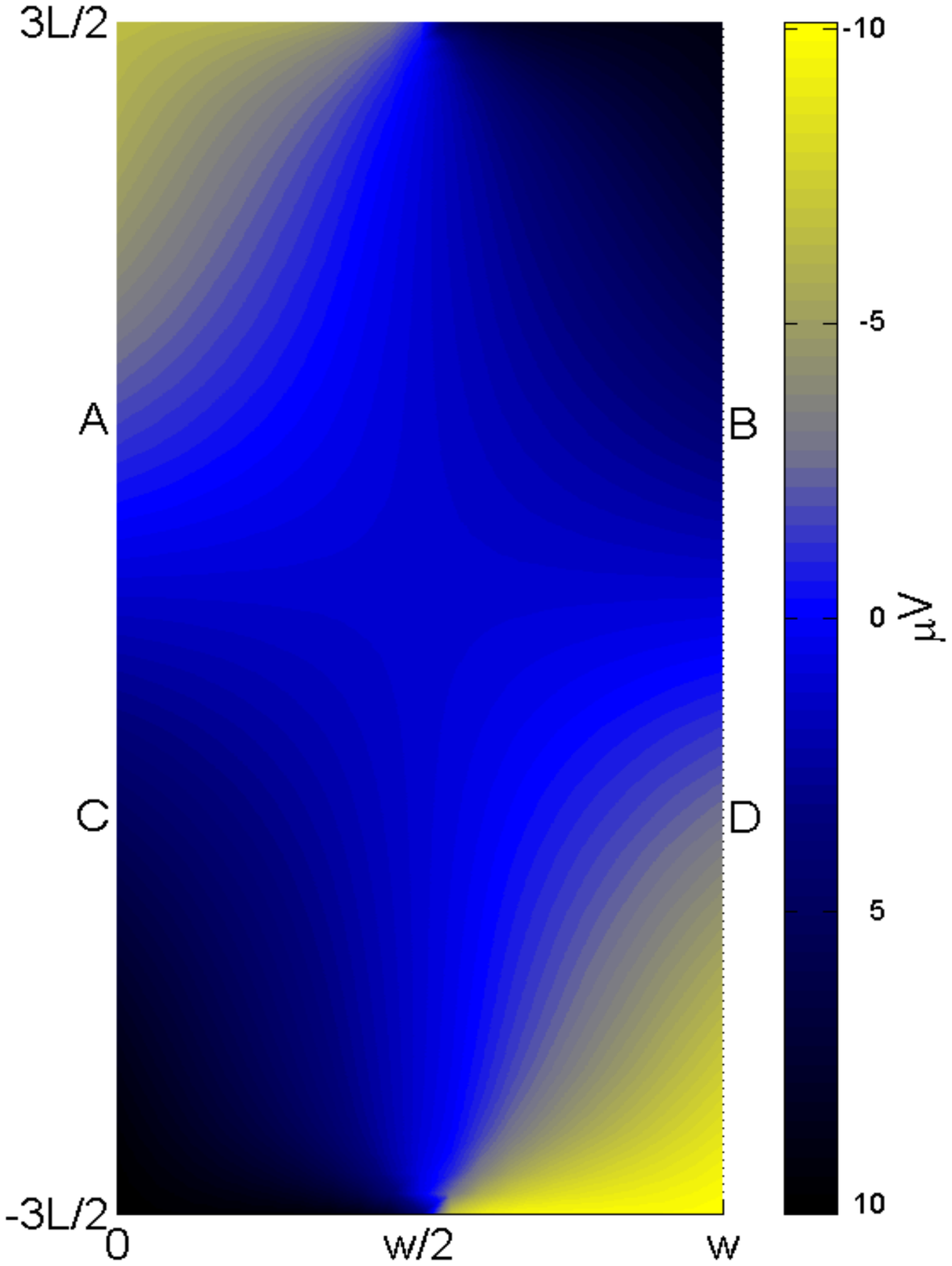}
\caption{\label{fig:five}Numerical calculation of electric potential generated by a nonuniform magnetization reversal.  The standard $I\cdot R$ contribution is subtracted for clarity. A,B,C and D correspond to locations of the voltage probes.}
\end{figure}


where $M_s$ is the saturation magnetization, $V_{AB,S}$ and $V_{CD,S}$ are the saturated values of $V_{AB}$ and $V_{CD}$ respectively. The normalized magnetization slope along the sample is then $\frac{\Delta M(H)}{L\cdot M_S}$. The solid line in Fig. \ref{fig:four} presents the simulation of the field dependent longitudinal voltage in the peak region for the Co/Pd multilayer sample with a single fitting parameter $\Delta t/t = 0.05$. The dashed line was calculated by Eq. \ref{eq:eight} and is shown here for comparison.  Numerical results agree nicely with the experimental data ($\circ$). A snapshot of the simulated sample potential during a gradual magnetization reversal (magnetization is zero at $x = 0$ at applied field of 0.44 T) is shown in Fig. \ref{fig:five}. Large Ohmic component $\frac {\rho (0)\cdot I \cdot x}{t\cdot w}$ was subtracted for clarity. It is clearly seen that the potential gradient along the sample has opposite polarities at two edges of the sample. It is important to note that due to the sharpness of magnetization reversal in films with an out-of-plane anisotropy a minor delay in coercive field (20 Oe as compared with 700 Oe of the reversal width) results in a relative difference of magnetization of up to about 20\% between cross- sections AB and CD, which respectively leads to distinctive voltage peaks in magnetoresistance.  

   To summarize, we studied the asymmetric field dependence of magnetoresistance in magnetic films. We argue that minor variation of thickness, Hall coefficient and nonuniform magnetization reversal along the sample can explain the anomalous phenomena. We show that a non-uniform variation of the Hall voltage along the sample generates an additional odd in field longitudinal voltage signal proportional to the field derivative of the transverse voltage. This additional signal can be significant when the Hall voltage varies sharply with the applied field, like in the case of magnetization reversal in films with perpendicular magnetic anisotropy, studied here, at superconducting transitions or in materials demonstrating the quantum Hall effect. The fingerprint of the mechanism is the reversal of the asymmetry when the longitudinal voltage is measured along the opposite edge of the sample.
   
   This work was supported by the Israel Science Foundation grant No. 633/06 and by the Air Force Office of Scientific Research, Air Force Material Command, USAF grant No. FA8655-07-1-3001.





\bibliography{citations}

\end{document}